\documentclass[12pt]{article}
\usepackage[cp1251]{inputenc}
\usepackage[english]{babel}
\begin{document}
\centerline{\Large\bf {Hilbert Transform: A New Integral Formula}}

\

\centerline{A. Alenitsyn, M. Arshad, A. S. Kondratyev, I.
Siddique}\vspace{2mm}

\centerline{\it School of Mathematical Sciences, GCU, Lahore,
Pakistan}

\centerline{alenitsyn@mail.ru }

\centerline{\ imransmsrazi@gmail.com}

\

\

\centerline{\bf Introduction}

\

It is shown in Quantum statistical mechanics [1] that the spectral
function of one-particle states in general case can be represented
in the form
$$A(\vec p,\omega)=\frac{\gamma(\vec p,\omega)}
{(\omega-\sigma(\vec p,\omega))^2+\frac14\,\gamma(\vec
p,\omega)^2},$$where real functions $\gamma$ and $\sigma$ are
related through Hilbert's transform [2]. The spectral function plays
central role in Quantum statistics. It was proved [1] that
$$\frac{1}{2\pi}\int\limits_{-\infty}^\infty A(\vec p,\omega)\,d\omega=1$$
for any $\vec p$.

Recently, H. S. K\"ohler [3] has stated a Conjecture: \ The
sufficient condition for the validity of the above equality is that
the functions $\gamma$ and $\sigma$ be related through the Hilbert
transform.

A proof of the Conjecture is not known to exist. In the present
paper we concern only pure mathematical aspects of the question. We
show that in general case the Conjecture is not valid: for a certain
set of functions $\gamma$ and $\sigma$, the equality holds while for
some others it fails.

\section{Formulation of the problem}

Let $f(x)$ be a real function defined for $x\in(-\infty,+\infty)$
and $g(x)$ be its Hilbert transform:
$$g(x)=\frac1\pi\,PV\int\limits_{-\infty}^{+\infty}\frac{f(t)\,dt}
{x-t}, \eqno(1)$$ where the integral is understood as Cauchy
principal value. Hilbert transform exists at least for continuous
functions tending rapidly to zero at infinity.

We define the {\it spectral function}, $S(x)$, by the formula
$$S(x)=\frac1\pi\cdot\frac{f(x)}{(x-g(x))^2+(f(x))^2}.\eqno(2)$$
It will be shown that the Conjecture in general case is not valid,
thus the next problem arises: \ Describe the subset of functions
$f(x)$ which satisfy the equality
$$\int\limits_{-\infty}^{+\infty}S(x)\,dx=1.\eqno(3)$$

\section{Exact results}

We offer three examples where the {\it spectral integral} \
$\int\limits_{-\infty}^{+\infty}S(x)dx=1$ \ can be calculated
explicitly. In first two cases the integral is exactly equal to 1
independently of the scale factor $\alpha$, while the third example
reveals the case when the integral is strictly less than 1.
\vspace{2mm}

{\it \textbf{Example 1}}. \ Let $f(x)=\alpha\sqrt{|x|}$ with a
positive constant $\alpha$. It is easy to show that
$$g(t)=-\alpha\sqrt{|t|}\,{\rm sign}\,(t),\ \ \ {\rm
sign}\,(t)=\cases{-1,\ t<0\cr\ 0,\ \ t=0\cr \ \ 1,\ t>0}.$$ The
spectral integral
$$\int\limits_{-\infty}^{+\infty}S(x)\,dx=
\frac\alpha\pi\int\limits_{-\infty}^{+\infty}\frac{\sqrt{|x|}\,dx}
{(|x|+\alpha\sqrt{|x|})^2+\alpha^2|x|}$$ by means of the
substitution $\sqrt{x}=\alpha y$ is reduced to \
$\frac4\pi\int\limits_0^{\infty}\frac{dt}{(y+1)^2+1},$ \ which
equals 1.

{\it \textbf{Example 2}}. \  Let $f(x)=\frac{\alpha}{x^2+1},\
\alpha>0$, then $g(t)=\frac{\alpha\,t}{t^2+1}.$ The spectral
integral can be simplified to \
$$\frac{1}{\pi}\int\limits_{-\infty}^{+\infty}
\frac{\alpha\,dx}{(x^2-\alpha)^2+x^2}$$ which can be evaluated in
term of residues. The result is exactly 1 for any $\alpha>0$.

{\it \textbf{Example 3}}. \  Let $f(x)=\frac{\alpha
x^2}{(x^2+1)^2},\ \alpha>0$, then \
$g(t)=\frac{\alpha\,t(t^2-1)}{2(t^2+1)^2}.$ \ The spectral integral
can be reduced to
$$\frac{4}{\pi}\int\limits_{-\infty}^{+\infty}
\frac{\alpha\,dx}{(2x^2+2-\alpha)^2+8\alpha}.$$ Using residues, we
get ultimately \ $\frac{\alpha}{2+\alpha}.$ \ We see that the
integral is strictly less than 1 for any $\alpha>0$.

\section{Numerical experiments}

We tested with the help of a computer the following functions
$f(x)$: \vspace{2mm}

1. \ Gaussian function \ $\alpha e^{-x^2/2}$; \vspace{2mm}

2. \ Valley function \ $\alpha (x^2+b)/(x^2+1)^2$;\vspace{2mm}

3. \ Rectangle function \ $\alpha\phi(x)$, where $\phi(x)=\cases{1,\
-1\le x\le 1\cr 0,\ |x|>1}$;\vspace{2mm}

4. \ Peak function \ $\alpha\psi(x)$, where $\psi(x)=\cases{1+x,\
-1\le x<0\cr1-x,\ 0\le x\le 1\cr 0,\ |x|>1}$.\vspace{2mm}

Hilbert transform for each of the above functions can be easily
found analytically. We computed the spectral integral using the
mathematical packages Maple and Derive; the computation was done
with the accuracy of 12--20 digits.

\subsection{Gaussian function}

If \ $f(x)=\alpha\,e^{-x^2/2}$, \ $\alpha>0$, \ \ then
$g(t)=\alpha\,\sqrt{\frac2\pi}\, e^{-x^2/2}\int\limits_0^x
e^{\,t^2/2}\,dt$.

Reliable results were obtained for values of $\alpha$ from 0.0001 to
40; the spectral integral proved to be equal to 1 with high
precision. For $\alpha$ greater than 40, the results became
unstable.

\subsection{Valley function}

If \ $f(x)=${\large$\frac{\alpha(x^2+b)}{(x^2+1)^2}$},\ $\alpha>0,\
b>0$, \ then \ $g(t)=${\large$\frac{\alpha\,t}{2(t^2+1)^2}$}
$(t^2(b+1)+3b-1)$\vspace{2mm}

We computed the spectral integral for $0.001\le b\le 10$ and
$0<\alpha<10000$; it appeared to be equal to 1 with high precision.

\subsection{Rectangle function}

If \ $f(x)=\cases{\alpha,\ -1\le x\le 1\cr 0,\ \ \ |x|>1}$, \ \ then
\ $g(t)=\,${\large$\frac\alpha\pi\ln\left|\frac{1+t}{1-t}\right|$}.

\

The computations show that the spectral integral $I(\alpha)$ is a
monotonously decreasing function. In particular, \vspace{2mm}

$I(0.00001)=0.999999680$, \ $I(0.1)=0.999476$, \ $I(0.5)=0.955$, \
$I(0.65)=0.8644$, \ $I(1)=0.6225,\ I(5)=0.1112,\ I(10)=0.0540$.

\subsection{Peak function}

If $f(x)=\cases{\alpha(1+x),\ -1\le x<0\cr\alpha(1-x),\ \ \ 0\le
x\le 1\cr 0,\ |x|>1}$\vspace{2mm}

\noindent then \ $g(t)=\,${\large$\frac\alpha\pi$}
$[(1+x)\ln|1+x|-(1-x)\ln|1-x|-x\ln(x^2)]$\vspace{2mm}

The equality (3) is valid not for all $\alpha>0$, in particular, for
$0<\alpha\le2.26$, it holds with high precision, but for
$\alpha=2.27,$ the equality suddenly fails: $I(2.27)=0.6945,\
I(3)=0.2799$ and so on.

\section{Conclusion}

It seems that the set of functions which contain the scale factor
$\alpha$ can be divided into two subsets: first one, where the
formula (3) holds with any value of $\alpha$, and the other one,
where the validity of (3) is conditioned by the value of $\alpha$.

It is interesting to notice that in most examples considered above
the graph of the spectral function $S(x)$ has, for small values of
$\alpha$, the form of a sharp peak centered around $x=0$, see the
Figure 1: it looks like the Dirac delta-function $\delta(x)$. The
parameter $\alpha$ being larger, the peak becomes lower and wider.
At some values of $\alpha$ it is transforming into two "hills"
located symmetrically with respect to the origin. When $\alpha$ is
increasing further, the two hills go farther from each other and
become more and more sharp like two delta-functions.

Another interesting observation concerns the function $f(x)$
represented as the sum of two nonnegative functions,
$f(x)=f_1(x)+f_2(x)$: if the spectral integral for each of the
summed equals 1. This property looks somewhat strange since the
spectral integral is an essentially nonlinear functional of $f(x)$.
Moreover, let a nonnegative function $f(x)$ give to the spectral
integral a value less than one, and $f_2(x)=\frac{\alpha}{x^2+1}$
with arbitrary small positive $\alpha$, then the sum $f_1(x)+f_2(x)$
provides that the spectral integral is again 1.

The observations described in the present paper can probably
stimulate the research of conditions sufficient for validity of the
formula (3).

\

\center \textbf{References}

\

[1] L. P. Kadanoff and G. Baym. Quantum statistical mechanics. W. A.
Benjamin, N.Y., 1962; Perseus Book, Cambridge, Massachusetts, 1989.

[2] CRC Concise Encyclopedia of Mathematics, by Eric W.Weisstein.
Chapman\&Hall/CRC, 2003.

[3] H. S. K\"ohler. Phys. Rev. C46, No 5, 1687 (1992).

\end{document}